\documentclass[amsmath,amssymb,pra,twocolumn,superscriptaddress,aps,10pt]{revtex4-2}
\usepackage{graphicx}
\usepackage{dcolumn}
\usepackage{bm}
\usepackage{booktabs}
\usepackage{gensymb} 
\usepackage{siunitx} 
\usepackage[colorlinks,linkcolor=blue,anchorcolor=blue,citecolor=blue,]{hyperref}
\usepackage{xcolor}

\usepackage{float}
\usepackage{booktabs}

\AtBeginDocument{%
    \heavyrulewidth=.08em
    \lightrulewidth=.05em
    \cmidrulewidth=.03em
    \belowrulesep=.65ex
    \belowbottomsep=0pt
    \aboverulesep=.4ex
    \abovetopsep=0pt
    \cmidrulesep=\doublerulesep
    \cmidrulekern=.5em
    \defaultaddspace=.5em
}

\begin{document}

\title{Optically Derived Radio-Frequency Benchmark in Methanol: A Sub-kHz Reference for Astrophysical Tests of Fundamental Physics}

\author{Frank M. J. Cozijn}
\affiliation{Institute of Condensed Matter and Nanosciences, Université catholique de Louvain, Louvain-la-Neuve, Belgium}
\affiliation{Department of Physics and Astronomy, LaserLab, Vrije Universiteit\\
De Boelelaan 1100, 1081 HZ Amsterdam, The Netherlands}

\author{Arthémise Altman}
\affiliation{Institute of Condensed Matter and Nanosciences, Université catholique de Louvain, Louvain-la-Neuve, Belgium}

\author{Alexandr S. Bogomolov}
\affiliation{Institute of Condensed Matter and Nanosciences, Université catholique de Louvain, Louvain-la-Neuve, Belgium}

\author{Alexis Libert}
\affiliation{Institute of Condensed Matter and Nanosciences, Université catholique de Louvain, Louvain-la-Neuve, Belgium}

\author{Simon Collignon}
\affiliation{Institute of Condensed Matter and Nanosciences, Université catholique de Louvain, Louvain-la-Neuve, Belgium}

\author{Erik Dierikx}
\affiliation{VSL Dutch Metrology Institute,
Thijsseweg 11, 2629 JA Delft, The Netherlands}

\author{Jeroen C. J. Koelemeij}
\affiliation{Department of Physics and Astronomy, LaserLab, Vrije Universiteit\\
De Boelelaan 1100, 1081 HZ Amsterdam, The Netherlands}

\author{Boy Lankhaar}
\affiliation{Institute of Theoretical Astrophysics, The Faculty of Mathematics and Natural Sciences, \\
University of Oslo, Sem Sælands vei 13, Oslo, Norway}

\author{Isabelle Kleiner}
\affiliation{
Université Paris Cité and Univ Paris Est Creteil, CNRS, LISA, F-75013 Paris, France}

\author{Clément Lauzin}
\affiliation{Institute of Condensed Matter and Nanosciences, Université catholique de Louvain, Louvain-la-Neuve, Belgium}

\author{Wim Ubachs}
\affiliation{Department of Physics and Astronomy, LaserLab, Vrije Universiteit\\
De Boelelaan 1100, 1081 HZ Amsterdam, The Netherlands}
\email{\authormark{*}w.m.g.ubachs@vu.nl} 

\begin{abstract}

Methanol radio lines observed in space provide sensitive probes of whether the proton-to-electron mass ratio has changed over cosmic time, but such tests require laboratory rest frequencies with very high accuracy. Here we determine the frequency of the astrophysically important 12.2 GHz $3_{-1}$E -- $2_{0}$E transition of CH$_3$OH by measuring near-infrared rovibrational transitions rather than the microwave line directly. Using wavelength-modulated NICE-OHMS locked to an ultra-stable optical frequency comb and referenced via a fiber link to a hydrogen-maser frequency standard, we measure Lamb-dip frequencies near 1.4 $\mu$m (216 THz) with 10 Hz statistical reproducibility and absolute uncertainties as low as 130 Hz. Pairs of optical transitions sharing common upper levels form a triangulation scheme that yields the ground-state rotational combination difference. We obtain 12 178 596.415(135) kHz, improving on earlier molecular-beam microwave spectroscopy by a factor of 20 and agreeing with a recent free-induction-decay measurement. This result establishes a sub-kHz laboratory benchmark for a key radio-astronomical methanol line and demonstrates that optical triangulation can be extended to non-chiral molecules with internal rotation.

\end{abstract}

\date{\today}

\maketitle

\section{Introduction}

In the recent decade cavity-enhanced absorption techniques have found widespread application in molecular precision spectroscopy~\cite{Gianfrani2024}.
The combination of cavity-enhancement with frequency comb lasers has merged the two ingredients desired in spectroscopy, sensitivity and calibration accuracy. This allows for highly accurate measurements of frequency positions of saturated absorption lines in various molecules.
A special variant of cavity-enhanced techniques is the noise-immune cavity-enhanced optical-heterodyne molecular spectroscopy (NICE-OHMS) approach, invented already decades ago~\cite{Ye1996} and further developed as an optical precision tool~\cite{Foltynowicz2008b}.
The excellent detection sensitivity of NICE-OHMS was recently demonstrated by measuring the Lamb dip of an extremely weak quadrupole overtone transition in the H$_2$ molecule~\cite{Cozijn2023}.
As for frequency precision a number of studies have been performed reaching kHz~\cite{Kassi2018,Aiello2022} and even sub-kHz absolute accuracies, from Lamb dip data~\cite{Wang2017,Tan2022}. 
By averaging over large data sets of Doppler-broadened lines Reed et al.~\cite{Reed2020} succeeded in extracting transition frequencies at a mean average as low as 212 Hz.

In the present study a frequency-comb locked NICE-OHMS spectrometer is used for ultra-precision measurements of near-infrared transitions in the methanol (CH$_3$OH) molecule. 
To obtain superior stability and absolute accuracy, the spectroscopy laser is locked to an in-house ultra-stable laser (USL) of Hz stability through an optical frequency comb (OFC) and referenced via a fiber network to a remote hydrogen maser located at the Dutch National Metrology Institute (VSL).
Such fiber links to national metrology centers for high-precision spectroscopy were previously demonstrated in the precision measurement of the 1S--2S transition in the H-atom~\cite{Matveev2013}, 
in the measurement of the 10~$\mu$m transition in SF$_6$~\cite{Shelkovnikov2008}, on mid-infrared methanol transitions~\cite{Santagata2019}, as well as a recent sub-ppt calibration of a line in HCOOH~\cite{Leuliet2026}.

Recently, for the case of the water molecule a spectroscopic network was built based on precision measurements of vibrational transitions~\cite{Tobias2020,Diouf2022,Tobias2024}
from which frequencies of important water maser lines~\cite{Ubachs2024} and of water lines protected for radio astronomy~\cite{Altman2025} could be determined at improved accuracies. 
The present study extends this concept of deriving a rotational splitting from vibrational Lamb dip experiments to methanol.
In this internal rotor molecule, symmetry properties and selection rules are such that the procedure is simpler and more direct. 
In methanol, it is possible, due to internal rotation and associated degeneracies, to connect three molecular levels via parity-changing electric dipole transitions in a triangle.
The situation of electric-dipole triangulation was previously postulated to occur for rotational transitions in chiral molecules~\cite{Hirota2012} 
and this idea found experimental realization in a number of studies~\cite{Patterson2015,Sun2023,Lee2024}.
Here we exploit the fact that such triangulation can also occur for non-chiral molecules that exhibit internal torsional motion in deriving highly accurate frequencies of radio lines from vibrational Lamb dip measurements.

In this context, precision metrology studies are performed on pairs of vibrational lines, matching the combination difference between $3_{-1}$E and $2_0$E levels in the vibrationless ground state, at sub-kHz precision.
A precision metrology study of this radio line at 12.2 GHz is motivated by the fact that it is a sensitive line for probing a possible variation of the proton-to-electron mass ratio on cosmological time and distance scales~\cite{Jansen2011,Levshakov2011}.
It was used in studies constraining the relative variation of this fundamental constant to the $10^{-7}$ level at look-back times of 7.4 billion years in cosmic history~\cite{Bagdonaite2013a,Kanekar2015}.

\section{Vibrational line selection for the metrology study}
\label{Line-select}

\begin{figure}[htbp]
\centering\includegraphics[width=0.9\linewidth]{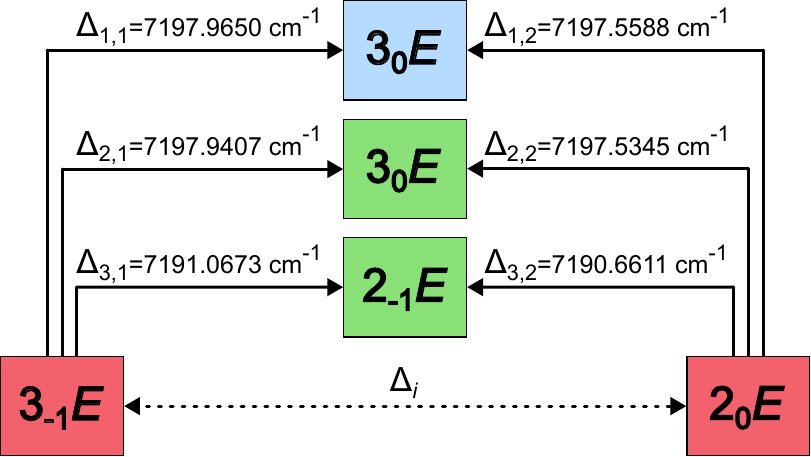}
\caption{Partial quantum level scheme of CH$_3$OH. Rotational states are indicated within colored boxes representing the vibrational ground (red) and excited (blue, green) levels. The green color-coded rotational levels pertain to the same excited vibration, whereas the blue color-coded rotational level refers to another excited vibration.
The six target transitions $\Delta_{i,j}$ (indicated with frequencies in cm$^{-1}$ units as obtained from the lower-resolution study~\cite{Libert2024}) form a triangulation scheme connecting three common upper levels ($i=1,2,3$) to the $3_{-1}$E ($j=1$) and $2_0$E ($j=2$) ground states to determine the splitting $\Delta_i$.
\label{fig:triangle}
}
\end{figure}

The goal of the present study is to determine the frequency separation between $3_{-1}$E and $2_0$E levels of CH$_3$OH in its vibrationless ground state through the measurement of vibrational transitions near $\lambda=1.4$ $\mu$m, the operation range of the NICE-OHMS setup~\cite{Tobias2020}.
While the rotational level structure of the ground vibration is accurately known~\cite{Lovas2004,Xu2008}, this does not hold for the vibrationally excited states. The region around 7200 cm$^{-1}$ is that of the ($2\nu_1$) overtone OH-stretching band as was investigated by pattern recognition methods~\cite{Rakovsky2021,Libert2024}, but the region is overlaid with other vibrational bands confusing the assignment of quantum numbers.
Although vibrational quantum numbers for these excited states cannot be unambiguously assigned, identification of the rotational quantum numbers remains feasible based on the well-established combination differences in the ground vibration. 
Before the present study, the separation between $3_{-1}$E and $2_0$E rotational levels in the ground state was known to a few kHz accuracy (the study of Ref.~\cite{Collignon2026} was performed concurrently alongside this work), and a number of paths in the vibrational structure could be identified that combine the two levels. This is shown in Fig.~\ref{fig:triangle}.

In atoms and simple diatomic molecules, a triangle connection as in Fig.~\ref{fig:triangle} would not be possible in view of changing parity in electric dipole transitions.
Hence, the two ground levels should have the same parity, prohibiting their connection via an electric-dipole allowed microwave transition.
For the same reason, the polyatomic H$_2^{~16}$O H$_2^{16}$O water molecule has four classes of quantum states related to para and ortho configurations and (+) and (-) parities in the (0,0,0) vibrational ground state.
These four classes cannot be directly connected via vibrational transitions~\cite{Tobias2020}.
The situation in molecules like methanol is different as will be explained in the next section.

\section{On the possibility of triangulation in methanol}
\label{sec:triangle}

The operation of the triangulation scheme employed in this work requires three energy levels to be mutually connected by allowed electric-dipole transitions. At first sight, such a closed loop appears to conflict with the usual parity selection rule. If the molecular eigenstates have well-defined parity, each electric-dipole transition connects states of opposite parity. Three successive electric-dipole transitions would therefore require an odd number of parity changes and cannot return to the parity of the initial state. In that simple situation a closed triangular transition scheme is forbidden.

The above argument relies on the relevant states being isolated parity eigenstates. This need not be the most useful description when symmetry-related states are effectively decoupled or degenerate. Chiral molecules provide a well-known example~\cite{Hirota2012}: although the full Hamiltonian is formally inversion-symmetric, the eigenstates corresponding to the two enantiomers are separated by a very large energy barrier and are therefore effectively decoupled. Once an enantiomer is formed, the molecule becomes trapped in that configuration, and its rotational eigenstates are no longer eigenstates of the inversion operator, thereby allowing the triangular transition scheme to operate~\cite{Quack1989}. Achiral symmetric-top molecules provide another important example. 
For $K\neq0$, the $K$ and $-K$ rotational states are degenerate and may be combined into states of opposite parity; this degeneracy likewise permits transition schemes that evade the simple parity argument~\cite{Hougen1962}.
Examples are found in the spectrum of phosphine~\cite{MalathyDevi2014,Ulenikov2021} and methyl bromide~\cite{Jacquemart2007}.

Methanol, the molecule of interest here, is an achiral asymmetric rotor, and therefore belongs to neither of the two cases just discussed. The way in which the parity argument is circumvented is instead tied to its non-rigidity: methanol exhibits hindered internal rotation (torsion) of the methyl group~\cite{Lin1959}. This internal motion has a profound impact on the spectrum and on the symmetry properties of the rotor eigenstates, which should properly be described as torsion–rotation eigenstates. When torsional motion is taken into account, the three protons of the methyl group become equivalent, and the appropriate symmetry description of methanol is given by the Longuet–Higgins permutation–inversion group $G_6$, which is isomorphic to $C_{3v}$ (see Refs.~\cite{Hougen1994,Hougen2001} and references cited therein).

Within this symmetry framework, the torsional eigenfunctions separate into A-type and E-type species, while the combined torsion–rotation eigenstates transform according to the irreducible representations $A_1$, $A_2$, or $E$. The A-type torsional functions couple straightforwardly to non-degenerate torsion–rotation eigenstates of $A_1$ or $A_2$ symmetry. In contrast, E-type torsional functions give rise to doubly degenerate torsion–rotation eigenstates. Crucially, these degenerate E-type states do not possess a well-defined parity but instead exhibit mixed parity~\cite{Hougen1994,Lankhaar2016}.
In a similar way as was discussed for the case of a symmetric top, the degeneracy of the E-type torsion–rotation states means that the levels involved in the transition loop cannot be assigned fixed parity labels. The simple parity argument against a closed electric-dipole loop therefore no longer applies.

Having established why the triangular scheme is allowed at the torsion–rotation level, we must still identify the states that are actually probed spectroscopically. These are the total hyperfine states obtained by coupling the torsion–rotation motion to the proton nuclear spins. Because the protons are fermions, Pauli’s exclusion principle imposes strict constraints on the symmetry of these total eigenstates, which must transform as either $A_1$ or $A_2$. For E-type torsion–rotation eigenfunctions, this requirement necessitates coupling to an E-type nuclear-spin eigenstate, yielding the direct product $E \times E = A_1 + A_2 + E$, of which only the $A_1$ and $A_2$ components are Pauli-allowed. As a consequence, a given E-type torsion–rotation eigenstate supports 8 total eigenstates, of which there are $4$ $A_1$ and $4$ $A_2$ symmetry eigenstates~\cite{Lankhaar2016}.

\section{METROLOGY DETAILS AND THE NICE-OHMS SETUP}

Vibrational overtone transitions in the methanol molecule at a wavelength of $1.4~\mu\text{m}$ were measured in a Noise-Immune Cavity-Enhanced Optical-Heterodyne Molecular Spectroscopy (NICE-OHMS) setup ~\cite{Cozijn-PHD,Tobias2020}. The principles of this technique, which combines a high-finesse optical cavity ($F \approx 150,000$) with frequency modulation spectroscopy (FMS) for near shot-noise-limited sensitivity, are well-documented~\cite{Ye1996,Foltynowicz2008b}.

In brief, an external cavity diode laser (ECDL) is phase-modulated by an electro-optic modulator (EOM) to create sidebands at the free-spectral-range (FSR) of the cavity, which for this setup is $404~\text{MHz}$. The optical resonator is arranged in a plano-concave configuration, employing a curved mirror with a radius of curvature (ROC) of $4000~\text{mm}$, yielding a mode beam waist of $0.72~\text{mm}$.
The laser carrier frequency is locked to a cavity mode using the Pound-Drever-Hall (PDH) technique. The molecular dispersion signal is detected in transmission and demodulated at the FSR frequency. A slow wavelength dither at $1.3~\text{kHz}$ is also applied to the cavity length, resulting in the technique of wavelength-modulated NICE-OHMS.
This allows for $1f$ lock-in detection to record a derivative of the dispersive lineshape, hence producing a symmetric lineshape. 
An acousto-optic modulator (AOM) is employed for power stabilization, utilizing the transmitted power through the cavity as a slow feedback loop to stabilize the intracavity roundtrip power over long timescales.

\begin{figure}[htbp]
\includegraphics[width=1\linewidth]{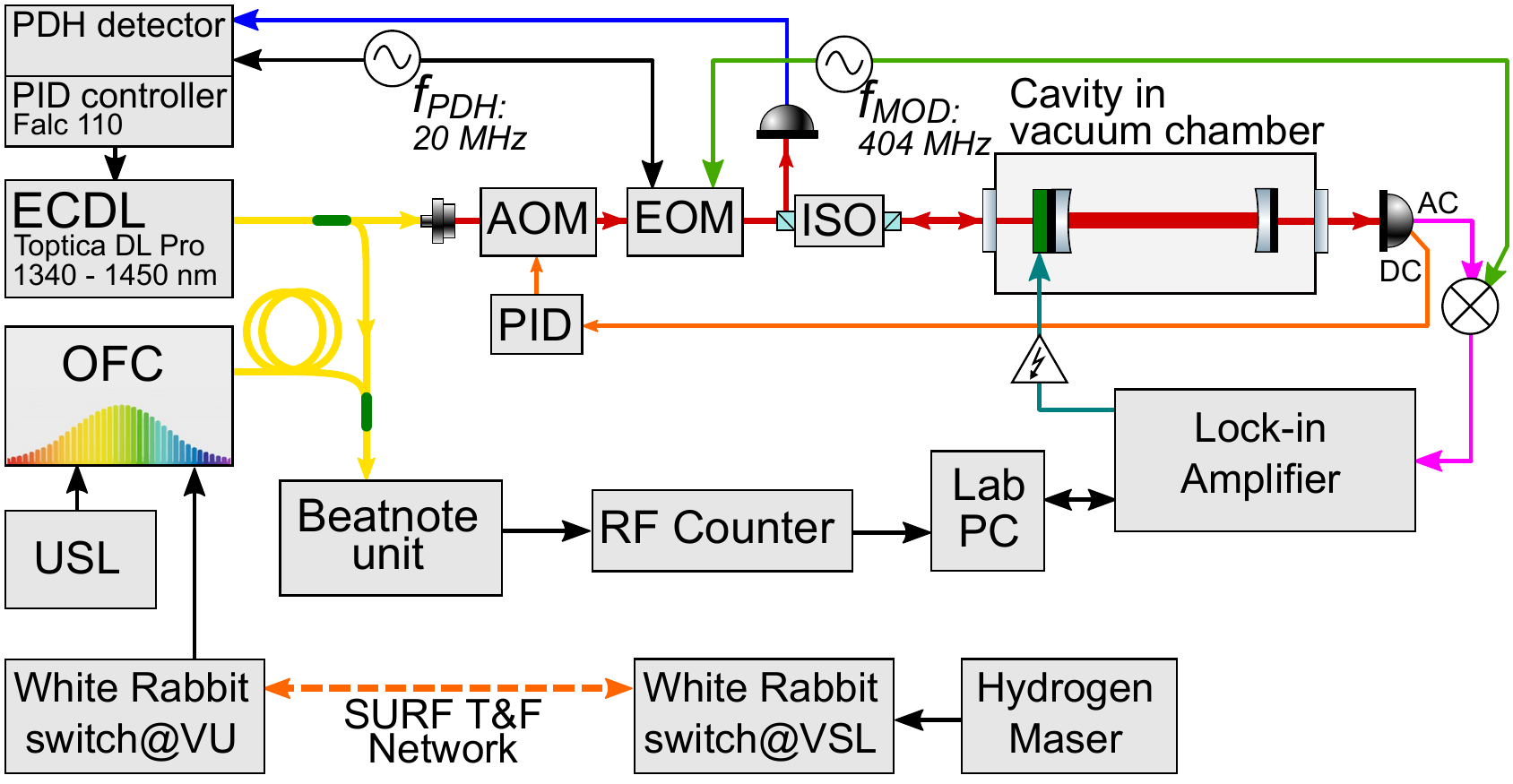}
\caption{Schematic of the experimental configuration including the details of the spectroscopy laser, the frequency comb laser, and the use of a frequency standard involving a fiber connection to UTC(VSL) at Delft, where UTC(VSL) is driven by a hydrogen maser. The (de)modulation and locking schemes are described in the main text.
\label{fig:setup}
}
\end{figure}

Stable sample pressure was maintained using a constant-flow system. To counteract pressure fluctuations from the tendency of methanol to adsorb to chamber surfaces, a dynamic equilibrium was established at typical pressures in the range 0.1--0.5 Pa, monitored by a capacitive pressure gauge (Pfeiffer CMR 375), by flowing methanol vapor from a liquid reservoir through a needle valve into the continuously pumped cavity.
All measurements were performed at {$-30^\circ\text{C}$ ($243~\text{K}$) by cooling the absorption cell in order to circumvent spurious water absorptions. 
The temperature was actively controlled to mK stability~\cite{Cozijn-PHD}.

Full technical details of the spectrometer, including a comprehensive analysis of systematic background drifts such as residual amplitude modulation (RAM), are provided in Ref.~\cite{Cozijn-PHD}. This prior characterization demonstrates near shot-noise-limited background stability for integration times up to 30 minutes. Because our individual scans are completed well within this timeframe, the instrumental baseline is entirely dominated by shot noise rather than potential systematic drifts.

The frequency accuracy for the experiment is established through a hierarchical stabilization scheme that anchors the spectroscopy laser to UTC(VSL), which is based on a steered active hydrogen maser (Safran iMaser 3000). Access to this standard is provided over a telecom fiber network through the SURF Time\&Frequency Network (TFN) using White Rabbit technology.

The OFC (Menlo FC1500-ULN nova) serves as the coherent frequency bridge for the experiment. Its carrier-envelope offset frequency ($f_0$) is phase-locked to a local Cs-clock (Microsemi CsIII Model 4310B), while the repetition rate ($f_\mathrm{rep}$) is stabilized against an in-house ultra-stable laser (USL; Menlo ORS) to yield Hz-level tooth stability. To ensure absolute calibration, $f_\mathrm{rep}$ is continuously counted against the reference frequency derived from UTC(VSL), provided over the TFN, allowing the absolute frequency to be reconstructed exactly at the time of measurement via post-processing.

\begin{figure}[htbp]
\includegraphics[width=1\linewidth]{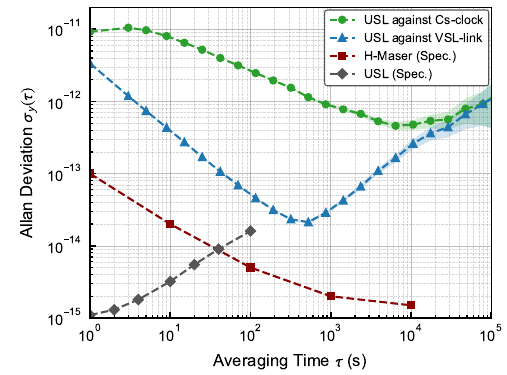}
\caption{Allan deviation $\sigma_y(\tau)$ representing the relative stability of the ultra-stable laser (USL) measured against (i) the local Cs-clock (green circles) and (ii) the hydrogen maser referenced via the White Rabbit fiber link (blue triangles). The optical frequency comb (OFC) serves purely as a coherent frequency bridge to translate the USL stability to the measured repetition rate ($f_{rep}$). Manufacturer specifications are included for the Hydrogen Maser (red squares) and the Ultra-Stable Laser (grey diamonds). The linear drift of the ultra-stable laser was removed in all instances. The TFN-link measurement (blue) reaches a stability floor of $\approx 2 \times 10^{-14}$, limited by fiber phase noise (short $\tau$) and residual non-linear laser drift (long $\tau$). The superior short-term stability of the ultra-stable laser (grey) relative to this floor validates its role as a flywheel; it bridges the gap dominated by fiber noise, effectively transferring the accuracy of the link-limited stability floor to the short timescales of the spectroscopic scan.
\label{fig:Allan}
}
\end{figure}

The improvement in absolute frequency stability provided by the remote hydrogen maser link is demonstrated in Fig.~\ref{fig:Allan}. Compared to the local Cs-clock reference, which requires nearly $10^4$ seconds to reach its optimal stability of $5 \times 10^{-13}$, the H-maser referenced system achieves this level in just 10 seconds. For both configurations, the Allan deviation displays a characteristic minimum (occurring at $\tau \approx 500$ s for the maser link and $\tau \approx 5000$ s for the Cs-clock) that marks the crossover point where the averaging of short-term reference noise is overtaken by the residual non-linear drift of the USL. Crucially, the superior short-term stability of the USL allows it to act as a flywheel that bridges these distinct timescales. Because the short-term stability of the USL remains well below the fiber link noise floor, it maintains phase coherence over the integration period. This flywheel mechanism effectively transfers the $\approx 2 \times 10^{-14}$ ultimate stability limit of the link-referenced standard directly to the 1-second timescale of the spectroscopic scan, ensuring that the individual comb teeth provide an ultra-stable, absolute optical grid. This grid acts as the frequency anchor for our laser system. To establish the absolute SI-traceability of this anchor, the stability of UTC(VSL) relative to true UTC was verified via the BIPM database for the exact window of the measurement campaign (MJD 60626 to 60637). During this interval, the maximum time drift was bounded at 0.5 ns/day, confirming that any systematic fractional frequency offset from true UTC was negligible at $\sim 6 \times 10^{-15}$.

With this frequency anchor established, the spectroscopy laser is locked to the high-finesse cavity for short-term performance, narrowing its linewidth to the sub-kHz domain. To ensure long-term absolute accuracy, the beatnote between this cavity-locked laser and the OFC is measured by a frequency counter, which provides real-time feedback to the cavity piezo to stabilize its physical length. This architecture provides precise scanning control and resolution, which is empirically verified by the tracking loop data exhibiting a standard deviation of 50 Hz or better per individual data point over a 5-second integration window. Consequently, while the underlying absolute optical anchor operates at the $10^{-14}$ level, the instantaneous optical frequency uncertainty per individual data point resides in the $\sim 10^{-13}$ domain. By averaging over the duration of a full spectroscopic scan comprising dozens of discrete points, the absolute frequency calibration of a measurement ultimately recovers the low $10^{-14}$ limit.

\begin{figure}[htbp]
\includegraphics[width=1\linewidth]{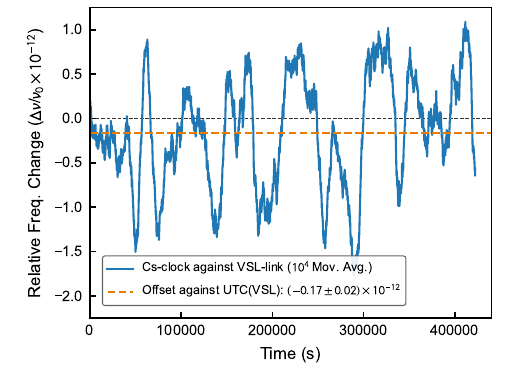}
\caption{
Frequency comparison of the in-house Cs-clock against the VSL Hydrogen maser, obtained by counting the Cs-clock 10 MHz signal against the H-maser over the TFN. The data (smoothed by a 10,000-second moving average) reveals fractional frequency deviations of $\pm 1 \times 10^{-12}$ and a systematic offset from UTC(VSL) (orange dashed line). Typically, UTC(VSL) drifts by less than 1 ns/day relative to UTC, corresponding to a negligible fractional frequency offset of $\lesssim 1 \times 10^{-14}$. The filtered time-domain data clearly illustrates the slow frequency wander, characteristic of flicker and random-walk noise, inherent to the free-running Cs-clock.
\label{fig:Cs_vs_maser}
}
\end{figure}

While the Allan deviation characterizes the stability of the Cs-clock reference in the variance domain, a time-domain representation better illustrates the practical impact of its long-term noise behavior during an extended measurement campaign. This is shown in Fig.~\ref{fig:Cs_vs_maser}, which plots the frequency evolution of the Cs-clock against the VSL H-maser reference received over the SURF TFN. The data is smoothed using a 10,000-second moving average. Acting as a low-pass filter, this moving average suppresses short-term white noise and effectively transmits frequency components slower than $(2\tau)^{-1}$, clearly revealing the underlying slow frequency wander. 

These significant frequency excursions of $\pm 1 \times 10^{-12}$ arise from the highly correlated stochastic noise, specifically flicker and random-walk frequency noise, that dominates the free-running Cs-clock at observation times $>5000$~s. For our optical transitions near $216~\text{THz}$, such an excursion introduces a time-varying systematic bias exceeding $200~\text{Hz}$. During parameter sweeps spanning several hours, this slow stochastic wander introduces varying absolute frequency offsets to sequential measurements (Fig.~\ref{fig:Cs_vs_maser}), exposing a critical limitation of the Cs-clock reference.

\section{Lamb dip Measurements of vibrational transitions}
\label{Lambdip}

To determine the $3_{-1}\mathrm{E}\text{--}2_0\mathrm{E}$ ground state combination difference, six target rovibrational transitions, grouped into three pairs, were measured as sub-Doppler Lamb dips using the NICE-OHMS setup. As schematically shown in Fig.~\ref{fig:triangle}, we denote these transitions $\Delta_{i,j}$. The index $i \in \{1,2,3\}$ identifies one of the three distinct excited rovibrational states that serve as a common upper level for a pair. The index $j \in \{1,2\}$ refers to the transition's lower level, either $3_{-1}$E ($j=1$) or $2_0$E ($j=2$). This scheme provides three independent measurements of the target ground state splitting, $\Delta_i$, by taking the frequency difference of each pair: $\Delta_i = \nu(\Delta_{i,1}) - \nu(\Delta_{i,2})$.

Individual scans were recorded over roughly 10-minute intervals. A typical scan comprised 50 discrete frequency points, with an active integration time of 5 seconds per point. These scans yielded signal-to-noise ratios (SNRs) of $\sim 1000$ at a circulating intracavity power of $\sim 1$~W.
A representative recorded spectrum is displayed in Fig.~\ref{fig:spec}(a). In the following subsections, several aspects of the lineshape analysis and data treatment are discussed, leading to a comprehensive assessment of the uncertainty budget.

\begin{figure}[htbp]
\centering\includegraphics[scale=1]{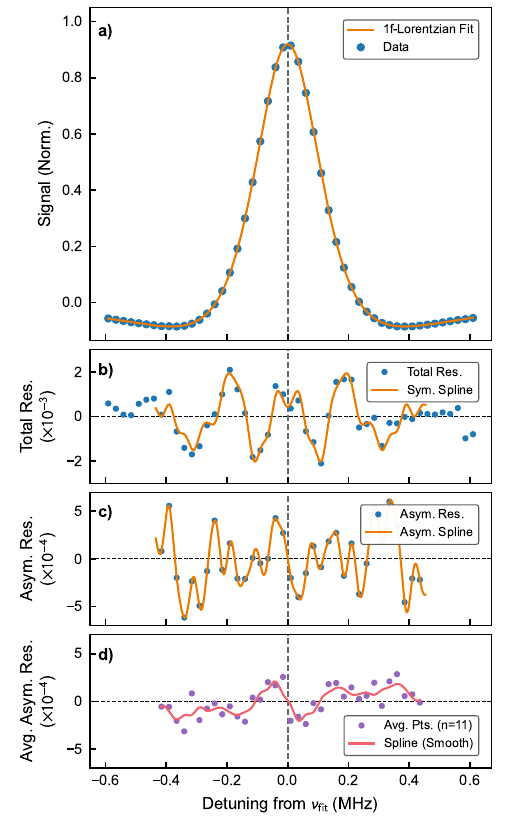}
\caption{A single scan of the $\Delta_{1,1}$ transition at 0.10 Pa pressure, 2 W intracavity roundtrip power and 150 kHz dither amplitude (peak-to-peak). (a) The recorded signal (points) fitted with the $1f$-dispersive Lorentzian profile (orange line), providing a 55 Hz fitting uncertainty. (b) Relative total residuals (points) from the fit, with the symmetric component (spline) shown in orange. The spline fit is truncated at the edges to avoid border defects. The (mostly) symmetric residuals are at a level of 0.4\% distortion. (c) Relative asymmetric residuals (points and spline), isolated by removing the symmetric component. These mostly show the randomly distributed noise, indicating a signal-to-noise level of around 3000. (d) Relative averaged asymmetric residual spectrum (points and smoothened spline) from 11 recordings, indicating an underlying asymmetry at the 0.04\% level.}
\label{fig:spec}
\end{figure}

\subsection{Lineshape fitting and systematic uncertainty}
\label{sec:lineshape}

A typical Lamb dip spectrum, illustrating the data quality and the fitting procedure, is shown in Fig.~\ref{fig:spec}(a). While the underlying NICE-OHMS signal provides a dispersive Lorentzian response, we implement an additional slow wavelength modulation (dither) to enable $1f$ lock-in detection. Consequently, the recorded lineshape corresponds to the first derivative of the dispersive profile.
The $1f$-dispersive Lorentzian fitting function used in our analysis is:
\begin{equation}
 S(\nu) = A \frac{\Gamma^2 - 4(\nu - \nu_{\mathrm{fit}})^2}{(\Gamma^2 + 4(\nu - \nu_{\mathrm{fit}})^2)^2} + B
 \label{eq:fit_func}
\end{equation}
where $S(\nu)$ is the demodulated signal as a function of frequency ($\nu$). The fit parameters are the line center ($\nu_{\mathrm{fit}}$), the amplitude ($A$), the linewidth parameter ($\Gamma$) in FWHM, and a constant vertical offset ($B$). For the representative scan displayed in Fig.~\ref{fig:spec}(a), the fitted linewidth $\Gamma$ amounts to 450~kHz. This observed width is a convolution of several mechanisms, primarily transit-time broadening, power broadening, modulation broadening, and the underlying, unresolved hyperfine structure. Collisional broadening is negligible at the sample pressure of 0.10~Pa.

To assess the quality of the fit and assign a realistic uncertainty to the extracted line center, we perform a detailed analysis of the residuals (data minus fit), shown in Fig.~\ref{fig:spec}(b). Notable symmetric structure is observed in the residuals. Attempting to account for this by using a Voigt profile in the fitting routine did not improve the fit quality, indicating the deformation is non-Gaussian. 
Furthermore, numerical simulations (detailed in the Supplementary Material) confirm that symmetric distortions arising purely from the unresolved underlying hyperfine structure, are over an order of magnitude smaller than the observed residuals. This indicates that the dominant symmetric deformation arises instead from instrumental origins such as modulation broadening~\cite{Supplee1994} or transit-time effects; the latter is known to induce non-Lorentzian lineshape distortions by favoring the detection of slow-moving molecules~\cite{Borde1976}. Crucially, any deviation in the residuals that is symmetric (even) with respect to the line center will not, to first order, pull or shift the fitted value of $\nu_{\mathrm{fit}}$.

However, any asymmetric (odd) component in the residuals indicates a distortion that directly biases the extracted line center. 
To separate and independently analyze both the symmetric and asymmetric deformations, we construct a continuous cubic spline interpolation of the total residuals. The symmetric spline shown in Fig.~\ref{fig:spec}(b) is the mathematically even component of these residuals evaluated around the fitted line center ($\nu_{\mathrm{fit}}$). Conversely, the asymmetric residuals in Fig.~\ref{fig:spec}(c) represent the odd component.

Quantifying the line-pulling effect induced by this asymmetry requires separating the persistent underlying structure from random statistical noise. To achieve this, we average the asymmetric residual spectra over multiple scans taken under identical experimental conditions, as shown in Fig.~\ref{fig:spec}(d). 
Speed-dependent collisional shifts (SDCS) were considered as a potential source of asymmetry, in view of theoretical models indicating such asymmetry scales with the magnitude of the total collisional shift~\cite{Wojtewicz2018}. However, given the low-pressure regime ($<0.3$ Pa) and small total shifts ($<3$ kHz), the induced asymmetry from SDCS is estimated to be negligible. Consequently, speed-dependent parameters were not included in the fit model. Instead, the observed residual asymmetry is attributed to a combination of instrumental imperfections and unresolved molecular structure. These primarily include dispersion-induced cavity effects, to which the NICE-OHMS technique is particularly susceptible, alongside minor contributions from modulation anharmonicity and unresolved hyperfine structure.

To estimate the lineshape asymmetry uncertainty, $\sigma\nu_{\text{asym}}$, induced by these unmodeled effects, we relate the magnitude of the averaged asymmetric component to the main signal amplitude using the relation:

\begin{equation}
 \sigma_{\text{asym}} = 2 \times \delta\nu_{\text{span}} \times \frac{S_{\text{pp,asym}}}{S_{\text{pp,signal}}}
 \label{eq:sys_err}
\end{equation}
Here, $S_{\text{pp,asym}}$ represents the peak-to-peak amplitude of the isolated asymmetric residual evaluated within the full-width at half-maximum ($\Gamma$) of the molecular transition, $S_{\text{pp,signal}}$ is the peak-to-peak amplitude of the measured lineshape, and $\delta\nu_{\text{span}}$ is the localized frequency span of the asymmetric features adjacent to the central zero-crossing. This relationship and empirical scaling factor of 2 is a conservative error envelope that safely bounds the worst-case line-pulling dynamics under all experimental conditions, validated through numerical lineshape simulations (see section III in the Supplementary Materials).

Applying this to a representative set of data pertaining to a point in the pressure-power parameter space for the $\Delta_{1,1}$ spectral line in Fig.~\ref{fig:spec}, 
it results in a lineshape asymmetry uncertainty contribution of approximately $62$~Hz. 
This magnitude indicates that the accuracy of individual scans is limited by lineshape systematics rather than the fitting uncertainty ($\sigma_{\text{fit}} \approx 55$~Hz).

\subsection{Statistical reproducibility}

The implementation of the SURF Time \& Frequency network link to UTC(VSL), which is driven by a hydrogen maser, provides a significant improvement in the statistical reproducibility of our measurements. This is demonstrated in Fig.~\ref{fig:rms}, which compares repeated measurements of a transition frequency referenced to either the H-maser or the in-house Cs-clock over a 2 hour measurement period. The standard error of the mean improved by more than a factor of two, decreasing from $32~\text{Hz}$ with the Cs-clock to $13~\text{Hz}$ with the H-maser. The Cs-clock referenced data not only exhibits a larger scatter but also a clear systematic offset from the H-maser referenced measurements. This discrepancy is a direct consequence of the observed frequency excursions documented in Fig.~\ref{fig:Cs_vs_maser}, which introduce a variable bias to the measurements.

\begin{figure}[htbp]
\centering\includegraphics[scale=1]{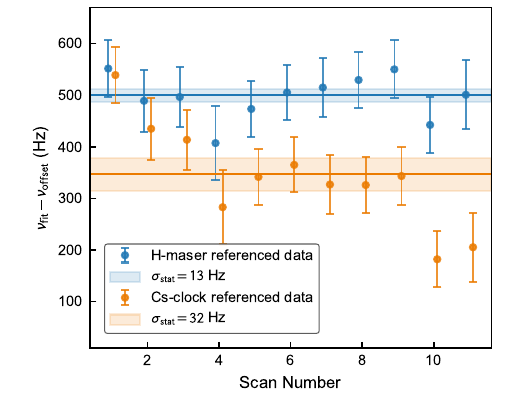}
\caption{
Statistical reproducibility of the $\Delta_{1,1}$ transition frequency ($\nu_{\mathrm{fit}}$) at $7197.9650~\mathrm{cm}^{-1}$. Data points represent individual scans recorded at 0.1~Pa pressure, 2~W intracavity power, and 150~kHz dither amplitude. The vertical axis displays $\nu_{\mathrm{fit}}$ relative to an offset $\nu_{\mathrm{offset}} = 215\,789\,561\,380~\mathrm{kHz}$. Error bars indicate the $1\sigma$ fit uncertainty ($\sigma_{\mathrm{fit}}$), while shaded bands represent the weighted standard error of the weighted mean ($\sigma_{\mathrm{stat}}$). The scans were referenced to either the fiber-linked hydrogen maser (blue) or the in-house Cs-clock (orange). Notable is the $153~\mathrm{Hz}$ shift between the weighted means, resulting in a discrepancy exceeding $4\sigma$ of the combined $\sigma_{\mathrm{stat}}$. This systematic shift is attributed to the non-statistical frequency wander of the Cs-clock (Fig.~\ref{fig:Cs_vs_maser}).
\label{fig:rms}
}
\end{figure}

\begin{figure*}[t!]
\centering
\includegraphics[width=\textwidth]{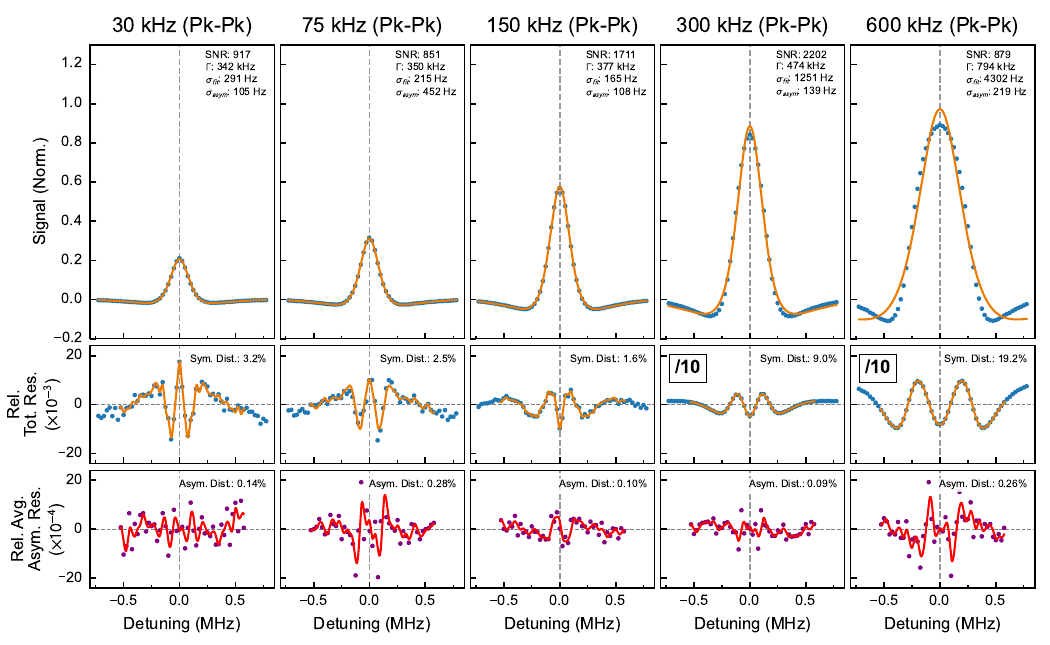}
\caption{Dither amplitude analysis for the $\Delta_{1,1}$ transition, performed at 0.12 Pa pressure and 1 W intracavity roundtrip power. The columns correspond to dither amplitudes ranging from 30~kHz to 600~kHz (peak-to-peak). 
\textbf{Top row:} Single-scan recordings of the acquired signal (blue points) with $1f$-dispersive Lorentzian fits (orange solid lines). 
\textbf{Middle row:} Relative total residuals (points) from the single scans shown above, with the isolated symmetric component (spline) overlaid as orange solid lines. Note that the residuals for 300~kHz and 600~kHz are scaled down by a factor of 10 (indicated by the boxed `/10' label) to accommodate the increased lineshape distortion. 
\textbf{Bottom row:} Relative averaged asymmetric residuals (points) averaged over $n=4$ consecutive scans, overlaid with a smoothed spline (red solid lines) to isolate line asymmetry. 
Signals in the top row are globally normalized to the 600~kHz peak to illustrate amplitude growth, whereas all residuals are normalized to their respective individual signal amplitudes.
The 150~kHz amplitude was selected as the optimal operating parameter, providing the lowest fitting uncertainty (165~Hz) and high SNR. 
Departures from this optimum result in reduced performance, limited by signal loss at lower dither amplitudes and by modulation broadening and distortion at higher amplitudes.}
\label{fig:dither_study}
\end{figure*}

To further validate this high level of precision and strictly rule out slow instrumental drifts, an extended reproducibility measurement comprising 60 consecutive scans over an 8-hour window was performed on the $\Delta_{3,1}$ transition. As detailed in the Supplementary Material, this analysis confirms the long-term stability of the extracted transition center and demonstrates the absence of any measurable systematic frequency drift from the apparatus. Crucially, this independent dataset yielded a commensurate statistical uncertainty of $\sim 10~\text{Hz}$, confirming our capacity to reliably and repeatedly access the 10~Hz precision domain. Thus, with the statistical uncertainty reduced to this level, the total uncertainty budget is no longer dominated by the frequency reference or statistical noise, but rather by the lineshape asymmetry effects discussed in the previous section. 

Consequently, for the subsequent 2D regression analysis (see section~\ref{sec:2D_regression}), the data is aggregated into discrete measurement sets, where each set comprises multiple independent scans recorded at a fixed pressure and optical power. For each aggregated set, we define a total uncertainty, $\sigma_{\text{tot}}$, as the quadrature sum of the statistical uncertainty and the estimated lineshape asymmetry component:
\begin{equation}
 \sigma_{\text{tot}} = \sqrt{\sigma_{\text{stat}}^2 + \sigma_{\text{asym}}^2}
 \label{eq:tot_error}
\end{equation}
Here, the statistical uncertainty ($\sigma_{\text{stat}}$) is calculated as the standard error of the mean of the fitted line centers within the measurement set, scaled by the appropriate Student's t-factor to account for the finite number of scans. In parallel, the lineshape asymmetry uncertainty ($\sigma_{\text{asym}}$) is evaluated for the entire set by averaging the asymmetric residuals of all constituent scans and applying Eq.~(\ref{eq:sys_err}). This combined $\sigma_{\text{tot}}$ represents the final uncertainty of the aggregated data point and is used as its statistical weight in the 2D regression (Sec.~\ref{sec:2D_regression}). Independent statistical weighting is justified since the lineshape asymmetry varies in both shape and sign across the pressure--power landscape.

\subsection{Dither amplitude selection}

The amplitude of the low-frequency ($1f$) wavelength modulation (at 1.3 kHz), or "dither," is a critical experimental parameter. For $1f$-detection of a dispersion profile, the demodulated signal is a pure first derivative of the lineshape only in the limit of an infinitesimal modulation amplitude. In practice, a finite amplitude is required to achieve a sufficient signal-to-noise ratio (SNR).

As the dither amplitude increases, the signal profile begins to include contributions from higher-order derivatives of the dispersion. 
This phenomenon, investigated previously by Supplee et al. \cite{Supplee1994}, and recently observed in Lamb dip resonances of water~\cite{Altman2025}, is responsible for the characteristic oscillating features observed in the fit residuals. 
While these contributions are theoretically symmetric, a large dither amplitude can non-linearly amplify any underlying experimental lineshape asymmetry (e.g., from unresolved hyperfine structure), potentially introducing a systematic bias in the fitted line center, $\nu_0$.

A dither analysis was therefore performed on the $\Delta_{1,1}$ transition to determine an optimal amplitude that maximizes the SNR (minimizing statistical uncertainty) while avoiding significant lineshape distortion (minimizing lineshape asymmetry). Based on the results shown in Fig.~\ref{fig:dither_study}, a dither amplitude of 150~kHz was selected as the best compromise for all subsequent measurements. At this amplitude a good signal strength is combined with the lowest fitting uncertainty.

To further investigate any potential systematic shift associated with this choice, the extracted line center was analyzed as a function of the dither amplitude. This analysis was performed for both the $\Delta_{1,1}$ and $\Delta_{1,2}$ transitions, with the results for $\Delta_{1,1}$ shown in Fig.~\ref{fig:dither_shift}. A weighted linear regression was applied to the data, yielding slopes of $-0.19 \pm 0.82$~Hz/kHz for $\Delta_{1,1}$ and $-0.17 \pm 1.87$~Hz/kHz for $\Delta_{1,2}$. In both cases, the determined slope is consistent with zero within its uncertainty, confirming that the line center determination is robust against the choice of dither amplitude and that no significant systematic bias is introduced.

\begin{figure}[htbp]
\includegraphics[width=1\linewidth]{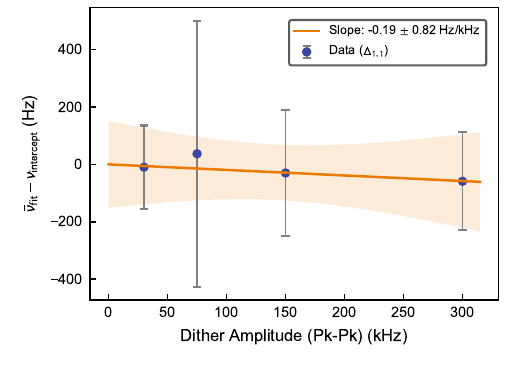}
\caption{
Based on the measurements recorded in Fig.~\ref{fig:dither_study}, this plot illustrates the dependence of the weighted averaged fitted line center ($\bar{\nu}_{\mathrm{fit}}$) of the $\Delta_{1,1}$ transition on dither amplitude. The vertical axis displays the frequency deviation $\bar{\nu}_{\mathrm{fit}} - \nu_{\mathrm{intercept}}$ (Hz). Error bars represent the total uncertainty ($\sigma_{\mathrm{tot}}$) for each dither setting, calculated as the quadrature sum of the statistical fit error and the lineshape asymmetry uncertainty. A weighted linear regression (orange line) yields a slope of $-0.19 \pm 0.82$~Hz/kHz. 
The slope's consistency with zero confirms that the line center extraction is robust against the choice of dither amplitude in the investigated range. A similar analysis for the $\Delta_{1,2}$ transition yields a slope of $-0.17 \pm 1.87$~Hz/kHz}, leading to the same conclusion.
\label{fig:dither_shift}
\end{figure}

\subsection{Neighboring resonance in $\Delta_{3,2}$}
\label{sec:neighboring_resonance}

\begin{figure}[htbp]
\includegraphics[width=1\linewidth]{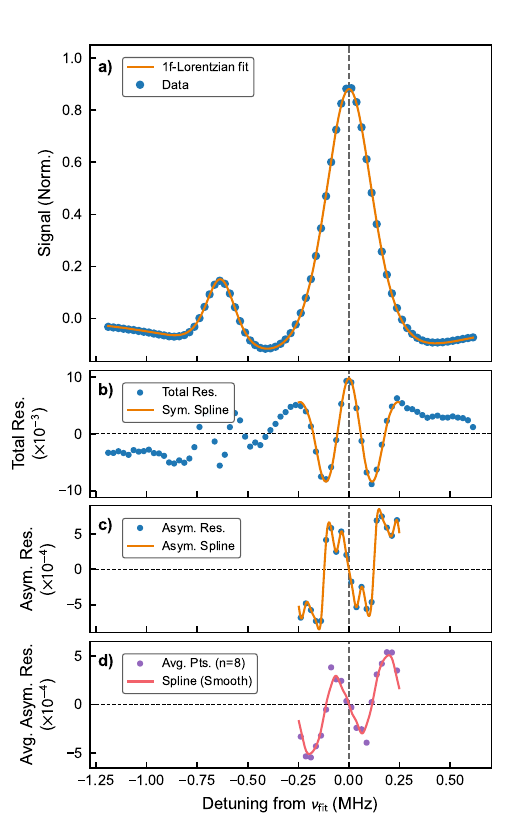}
\caption{Analysis of the $\Delta_{3,2}$ transition, complicated by a neighboring resonance. The panel layout and residual isolation methodology are identical to those detailed in Fig.~\ref{fig:spec}. (a) The recorded signal (blue points) is fitted with a double $1f$-Lorentzian profile (orange line) to account for the weaker, neighboring resonance at -600 kHz detuning. (b) The total residuals show a large, structured deviation due to the influence of the second peak; the symmetric spline fit is restricted to the range of the linewidth. (c, d) The single-scan and averaged asymmetric residuals are significantly larger than for isolated transitions, quantifying a persistent, non-statistical lineshape distortion that dominates the final total uncertainty for this transition.
\label{fig:delta32}
}
\end{figure}

A special case is that of the $\Delta_{3,2}$ transition, where the spectrum is complicated by the presence of a second, weaker resonance at a detuning of approximately -600~kHz, as shown in Fig.~\ref{fig:delta32}. 
While the two features, both measured as a Lamb dip, are sufficiently separated to be resolved, the spectral wing of the neighboring line introduces a significant, structured asymmetry to the lineshape of the target transition. This underlying asymmetry, clearly visible in the decomposed residuals (Fig.~\ref{fig:delta32}(d)), hinders an unambiguous extraction of the line center and significantly increases the fitting uncertainty. To maintain a consistent analysis, we account for this effect through our residual analysis framework, which leads to a larger final uncertainty for the $\Delta_3$ combination difference, reflecting the lower intrinsic quality of this particular measurement.

\subsection{2D regression analysis and systematic shifts}
\label{sec:2D_regression}

To determine the unperturbed transition frequencies, we must account for systematic shifts induced by sample pressure (collisional shifts) and intracavity power (AC Stark and saturation effects). 
While pressure shifts are typically dominant, power-dependent effects cannot be neglected at the precision level of this experiment. A distinct one-dimensional analysis of these parameters was hindered by the experimental difficulty in maintaining strictly constant pressure levels while varying the optical power. Consequently, to rigorously decouple these effects and extract the unperturbed frequency, we applied a simultaneous two-dimensional (2D) regression to the complete dataset.

We model the measured transition frequency $\nu(P, W)$ as a linear plane:
\begin{equation}
 \nu(P, W) = \nu(0,0) + c_p P + c_w W
 \label{eq:2d_fit}
\end{equation}
where $c_p$ is the linear pressure-shift coefficient (in \si{kHz/Pa}) and $c_w$ is the linear power-shift coefficient (in \si{kHz/W}). The intercept, $\nu(0,0)$, represents the extrapolated frequency at the zero-pressure, zero-power limit at $T \approx 243$~K.

The fit is performed using a multivariate Weighted Least Squares (WLS) regression on the complete $(\nu_i, P_i, W_i)$ dataset for each transition. To ensure the statistical rigor of the final extrapolated intercepts, extensive datasets were acquired. Depending on the transition, between 26 and 86 independent scans were recorded across the pressure-power parameter space, comprising a total of 369 scans and over 25 hours of active laser integration. The individual data points are then weighted by the inverse square of their total uncertainty ($\sigma_{\mathrm{tot}}$) derived from the lineshape analysis (Sec.~\ref{sec:lineshape}). To account for independent variable errors, the actual derived pressure variation during the scans is propagated into these fit weights using an iterative effective variance weighting until convergence is achieved. This ensures that the uncertainty of the final intercept is statistically rigorous.

\begin{figure}[htbp]
\includegraphics[width=1\linewidth]{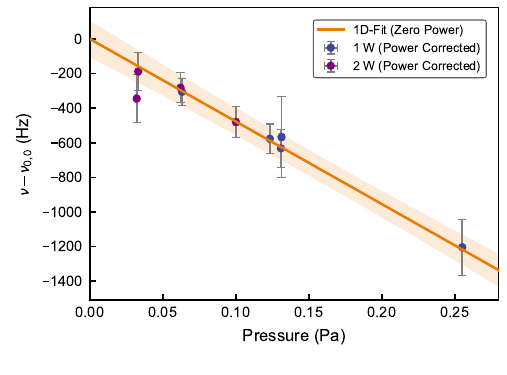}
\caption{
Power-corrected pressure dependence of the $\Delta_{1,1}$ transition. The vertical axis displays the frequency relative to the global fit intercept $\nu_{\mathrm{0,0}} = 215\,789\,561\,381\,282 \pm 107~\mathrm{Hz}$. Data points (blue and purple indicating different intracavity power levels) are projected onto the zero-power plane by subtracting the fitted power-dependent shift ($c_w W$) to isolate the collisional contribution. The vertical error bars include the uncertainty of this applied power correction, while the small horizontal error bars represent the derived sample pressure uncertainty. The solid line represents the zero-power projection of the global 2D fit ($c_p = -4.7 \pm 0.5~\mathrm{kHz/Pa}$). The orange band indicates the resulting $1\sigma$ confidence interval of this projection.}
\label{fig:pressure_shift}
\end{figure}

The results of this global analysis are summarized in Table~\ref{tab:final_frequencies}. To visualize the isolated pressure shift and highlight the quality of the data, Fig.~\ref{fig:pressure_shift} displays the frequency measurements for the $\Delta_{1,1}$ transition after correcting for the fitted power-induced shift ($c_w W$). The clear linear trend confirms the validity of the model.

The extracted coefficients reveal a significant state-dependence in the collisional perturbation. As shown in Table~\ref{tab:final_frequencies}, the $\Delta_{1}$ pair exhibits negative pressure shifts ($c_p \approx -5$ to $-11$~kHz/Pa), whereas the $\Delta_{2}$ pair shows positive shifts ($c_p \approx +5$~kHz/Pa). This distinct sign reversal highlights the state-dependent nature of the collisional physics in methanol, a behavior similarly observed in previous molecular Lamb dip experiments~\cite{Tobias2020,Diouf2021,Wang2025}.

Regarding power shifts, the coefficients $c_w$ are generally small, effectively correcting for minor AC Stark or saturation systematics. An exception is the $\Delta_{3,2}$ line, which exhibits a significantly larger coefficient ($c_w = 2.20$~kHz/W). As discussed in Sec.~\ref{sec:neighboring_resonance}, this is attributed to spectral contamination from the neighboring resonance, which induces a power-dependent lineshape distortion that the regression absorbs into the $c_w$ term.

The validity of the linear 2D model is confirmed by an analysis of the weighted residuals, which are found to be structureless across all datasets. Notably, nearly all residuals fall within $1\sigma$ of the fitted planes, with only a single data point marginally exceeding this threshold at $1.1\sigma$. Furthermore, $82\%$ of the residuals lie within $0.50\sigma$. This low scatter indicates that the short-term measurement reproducibility is better than the conservative uncertainty estimates assigned to individual points.

\subsection{Relativistic and Recoil Effects}
\label{sec:relativistic}

To account for small systematic frequency shifts, we consider the full relation between the measured photon energy and the unperturbed quantum level separation, $h\nu_0$~\cite{Cozijn2023}:
\begin{equation}
 h\nu(0,0) = h\nu_0 \pm \frac{h \vec{k}\cdot\vec{v}}{2\pi} - \frac{(h\nu_0)v^2}{2c^2} \pm \frac{(h\nu_0)^2}{2mc^2}.
 \label{eq:trans}
\end{equation}
The term with $\vec{k} \cdot \vec{v}$ represents the first-order Doppler effect, which is eliminated in our experiment by the nature of sub-Doppler Lamb-dip spectroscopy. The last two terms, representing the second-order Doppler shift and the photon recoil effect, are small but relevant at our level of precision and will be discussed.

\subsubsection{Second-order Doppler Shift}
The third term in Eq.~(\ref{eq:trans}) represents the second-order Doppler shift (SODS), a relativistic time dilation effect arising from the thermal motion of the molecules. The magnitude of this frequency shift, $\delta\nu_{\text{SODS}}$, is given by:
\begin{equation}
 \delta\nu_{\text{SODS}} = \nu_0 \frac{\langle v^2 \rangle}{2c^2}
 \label{eq:sods}
\end{equation}
where $\nu_0$ is the unperturbed transition frequency and $\langle v^2 \rangle$ is the mean-squared velocity of the molecules contributing to the saturation signal. 

In Lamb dip spectroscopy, the signal is generated selectively by molecules with zero longitudinal velocity ($v_z \approx 0$). Consequently, the Doppler shift is dominated by the thermal motion in the two transverse degrees of freedom, for which the mean-squared velocity is $\langle v^2 \rangle_{2D} = 2k_B T/m$. 
In the strict low-power limit, transit-time effects bias the effective velocity distribution toward slow-moving molecules, significantly reducing the SODS \cite{Hall1976b,Cozijn2023}. 
However, our measurements operate well into the saturated regime ($S_0 \ge 2.3$, see Supplementary Material)
This strong driving field heavily suppresses transit-time velocity selection, maintaining an effective velocity distribution very close to the standard thermal 2D average. Furthermore, because our 2D regression model relies on a linear extrapolation from this saturated data, the zero-power intercepts inherently retain this thermal velocity baseline.

For the present measurements at $T \approx 243$~K, the thermal 2D limit yields a shift of $\approx 151$~Hz. 
We apply this full value as the correction. To conservatively account for any minor residual velocity selection effects present at $S_0 \approx 2.3$, we assign a systematic uncertainty of 20\% ($30$~Hz).

\subsubsection{Recoil Effect}

The final term in Eq.~(\ref{eq:trans}) describes the photon recoil effect, which arises from the conservation of momentum during the absorption and emission of a photon. In case of full saturation of the transition splits into a doublet with components shifted by $\pm \delta\nu_{\text{recoil}}$, where $\delta\nu_{\text{recoil}} = h\nu_0^2/(2mc^2)$~\cite{Hall1976b}. For the near-infrared vibrational transitions in this study, the recoil splitting amounts to $\pm 3.2$~kHz, which remains unresolved. 
This contrasts the recent observation of a very weakly saturated H$_2$ quadrupole transition, where only the blue-shifted recoil component was observed~\cite{Cozijn2023,Ubachs2025}. 
However, in view of full saturation achieved in the present experiments, with a saturation parameter $S_0 \geq 2.3$ no net frequency shift occurs, and no correction for the recoil effect is imposed.

\subsection{Hyperfine centroid shift}

The potential shift of the centroid frequency resulting from the hyperfine substructure can be quantified, following a procedure described in Refs.~\cite{Kassi2022,Jozwiak2023,Collignon2026}. 
The $3_{-1}$E -- $2_{0}$E microwave transition at 12.2 GHz is used as a representative example, for which computed intensities of 22 hyperfine subcomponents~\cite{Lankhaar2016} were folded to simulate the experimentally observed linewidth.
The relative frequencies of the hyperfine lines with respect to the purely rovibrational center, as well as the Einstein $A$-coefficients are listed in the Supplementary Material.
The analytical first derivative of the composite dispersive lineshape, representing the zero-dither limit, is implemented for the hyperfine sub-components using line strengths $S\mu_i^2$, taken as proportional to the listed $A$ values and frequencies $\nu_i$. 
A theoretical centroid frequency can be calculated via:
\begin{equation}\label{eq:centroid}
	\nu_{\text{centroid}} =  \frac{\sum_{i} S\mu_i^2 \nu_i}{\sum_{i} S\mu_i^2}
\end{equation}

In case of underlying hyperfine components exhibiting a profile $\phi$ and a width $W$, a composite line shape $\phi^W_{\text{comp}}$ can then be computed:
\begin{equation}\label{eq:W}
	\phi^W_{\text{comp}}(\nu) =  \frac{\sum_{i} S\mu_i^2 \phi^W_i(\nu)}{\sum_{i} S\mu_i^2}
\end{equation}
This procedure can be performed for a width of the hyperfine components, producing a composite line of width $\Gamma$.} 
It can be shown that for this typical set of hyperfine components spanning a frequency range of $< 40$ kHz, the composite line becomes close to symmetric for $\Gamma > 200$ kHz.
Resulting values for the shift of the fitted line center (the experimental centroid) with respect to the theoretical centroid are plotted in Fig.~\ref{fig:niceohms_error}. 
For a representative narrow experimental width of $\Gamma=300$ kHz this shift is found to be 7 Hz. 
The magnitude of the hyperfine splittings in the excited vibrational states are not experimentally determined, but in good approximation the hyperfine constants do not differ over vibrational levels. Therefore this 7 Hz shift is also adopted as a centroid-related systematic uncertainty for the optical rovibrational transitions measured in this work.

\begin{figure}[!ht]
\centering
\includegraphics[width=.45\textwidth]{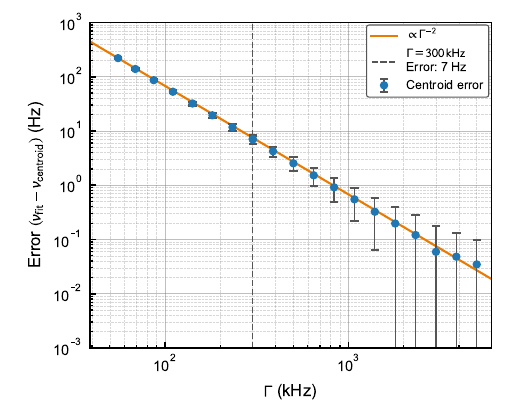}
\caption{Systematic shift between the fitted center frequency $\nu_{\mathrm{fit}}$ and the central centroid $\nu_{\mathrm{centroid}}$ for the $3_{-1}\mathrm{E} - 2_{0}\mathrm{E}$ transition. The tracking errors are extracted by fitting a single analytical $1f$ dispersive Lorentzian profile to the pure first derivative of the composite dispersive hyperfine multiplet. The shift converges rapidly to zero following a $\propto \Gamma^{-2}$ scaling law (orange line). For a representative narrow linewidth of $\Gamma = 300$~kHz, a systematic line-pulling offset of only 7~Hz is found; further details on the zero-dither and finite-dither simulation protocols are provided in the Supplementary Materials.}
\label{fig:niceohms_error}
\end{figure}

In the Supplementary Material a generalized analysis is added for centroid shifts pertaining to underlying substructure in optical transitions sharing a common level.

\subsection{Resulting Transition Frequencies and Error Budget}
\label{sec:results}

\begin{table*}[htbp]
\caption{Final results of the 2D Weighted Least Squares (WLS) regression and the final, corrected molecular frequencies. The extrapolated frequency $\nu(0,0)$ is the intercept at zero pressure and zero power (Sec.~V.G). $\sigma_{\nu(0,0)}$ is the $1\sigma$ standard error of this extrapolated intercept, propagating the total uncertainties of the individual data points. The pressure ($c_p$) and power ($c_w$) shift coefficients are shown with their $1\sigma$ fitting uncertainties. 
The final, at-rest molecular frequency, $\nu_0$, is determined by correcting the intercept for the Second-Order Doppler Shift magnitude ($\delta\nu_{\text{SODS}} = 0.151$~kHz) using the relation in the column header (Sec.~\ref{sec:relativistic}, \ref{sec:results}). 
The final uncertainty $\sigma_{\nu_0}$ is determined by linearly adding the total systematic error ($\sigma_{\text{sys,}\nu} = 0.031$~kHz) to the extrapolation uncertainty $\sigma_{\nu(0,0)}$. The $\sigma_{\text{sys,}\nu}$ systematic of 31 Hz is added linearly. All units are in kHz.}
\label{tab:final_frequencies}
\centering
\begin{ruledtabular}
\begin{tabular}{l c c r @{ $\pm$ } l r @{ $\pm$ } l c c}
Line & Extrapolated Intercept & Uncertainty & \multicolumn{2}{c}{Pressure Shift} & \multicolumn{2}{c}{Power Shift} & Transition Frequency & Final Uncertainty \\
($\Delta_{i,j}$) & $\nu(0,0)$  & $\sigma_{\nu(0,0)}$  & \multicolumn{2}{c}{$c_p$}  & \multicolumn{2}{c}{$c_w$}  & $\nu_0 = \nu(0,0) + \delta\nu_{\text{SODS}}$  & $\sigma_{\nu_0} =  \sigma_{\nu(0,0)} + \sigma_{\text{sys,}\nu}$  \\
 &  (\si{kHz})  & (\si{kHz}) & \multicolumn{2}{c}{(\si{kHz/Pa})} &  \multicolumn{2}{c}{(\si{kHz/W})} &  (\si{kHz}) &  (\si{kHz}) \\
\midrule
$\Delta_{1,1}$ & \num{215789561381.282} & 0.107 & $-4.7$ & $0.5$ & $-0.15$ & $0.05$ & \num{215789561381.433} & 0.138 \\
$\Delta_{1,2}$ & \num{215777382784.960} & 0.216 & $-11.2$ & $0.9$ & $-0.02$ & $0.10$ & \num{215777382785.111} & 0.247 \\
\addlinespace
$\Delta_{2,1}$ & \num{215788833269.074} & 0.109 & $5.4$ & $0.5$ & $-0.20$ & $0.15$ & \num{215788833269.225} & 0.140 \\
$\Delta_{2,2}$ & \num{215776654672.639} & 0.098 & $5.9$ & $0.4$ & $0.43$ & $0.18$ & \num{215776654672.790} & 0.129 \\
\addlinespace
$\Delta_{3,1}$ & \num{215582773999.670} & 0.677 & $13.3$ & $0.7$ & $-0.12$ & $0.99$ & \num{215582773999.821} & 0.708 \\
$\Delta_{3,2}$ & \num{215570595402.191} & 1.059 & $-3.5$ & $4.4$ & $2.20$ & $0.91$ & \num{215570595402.342} & 1.090 \\
\end{tabular}
\end{ruledtabular}
\end{table*}

The values for $\nu(0,0)$ presented in Table~\ref{tab:final_frequencies} represent the unperturbed experimental frequencies, extrapolated to zero pressure and zero power at the set temperature ($T \approx 243$~K). Table~\ref{tab:final_frequencies} consolidates the full analysis chain, presenting these extrapolated intercepts along with the shift coefficients ($c_p$, $c_w$) derived from the 2D regression (Sec.~\ref{sec:2D_regression}). Note that the effects of localized lineshape asymmetry ($\sigma_{\text{asym}}$) are already incorporated directly within this regression model as point-by-point statistical weights, establishing the baseline statistical extrapolation error, $\sigma_{\nu(0,0)}$.
Final, at-rest molecular transition frequencies, $\nu_0$, are then determined by correcting for the second-order Doppler shift (SODS), of $\delta\nu_{\text{SODS}} \approx 151$~Hz. 

The final uncertainty budget id mainly governed by the statistical analysis. The SODS correction carries an intrinsic uncertainty of $30$~Hz, the centroid shift related to hyperfine structure adds 7~Hz. 
The absolute tracking uncertainty from the remote hydrogen maser fiber link is negligible at $<1$~Hz.
The independent systematics are taken in quadrature of $\sigma_{\text{sys,}\nu}$ yielding 31~Hz, which is then added linearly to the statistical extrapolation error ($\sigma_{\nu(0,0)}$) to determine the final, absolute uncertainty of each rovibrational line.

\section{Extraction of rotational radio line frequency}
\label{sec:comb_diff}

\begin{table}[htbp]
\caption{Resulting extraction of the frequency for the $3_{-1}$E -- $2_0$E radio line. The $\Delta_i$ values are the three independent determinations derived from the combination differences in this work. 
Their purely statistical uncertainties ($\sigma_{\text{stat}}$) are calculated as the quadrature sum of the statistical extrapolation uncertainties ($\sigma_{\nu(0,0)}$) of the two constituent optical transitions. Because the 7~Hz hyperfine shift acts independently on each optical line, it contributes a combined systematic uncertainty of $\sigma_{\text{sys,}\Delta} \approx 10$~Hz to the difference. This is added linearly to $\sigma_{\text{stat}}$ to yield the total uncertainty ($\sigma_{\Delta}$). The ``Final result'' row represents the final, aggregated frequency of this experiment, calculated using purely statistical weighting (based solely on $\sigma_{\text{stat}}$). Its uncertainty reflects the statistical standard error of the weighted mean plus the irreducible 10~Hz systematic boundary. Previous literature values from direct microwave measurements are included for comparison. All units are in kHz.}
  \label{tab:radio}
  \centering
\begin{tabular}{c c c c c}
\toprule
Triangle & Frequency & \multicolumn{3}{c}{Uncertainty (\si{kHz})} \\
\cmidrule(lr){3-5}
 & (\si{kHz}) & $\sigma_{\text{stat}}$ & $\sigma_{\text{sys,}\Delta}$ & Total ($\sigma_{\Delta}$) \\
\midrule
$\Delta_{1}$  &  12\,178\,596.322 &  0.241 & 0.010 & 0.251  \\
$\Delta_{2}$  &  12\,178\,596.435 &  0.147 & 0.010 & 0.157  \\
$\Delta_{3}$  &  12\,178\,597.479 &  1.257 & 0.010 & 1.267 \\
\midrule
 \textbf{Final result}   &  \textbf{12\,178\,596.415} & \textbf{0.125} & \textbf{0.010} & \textbf{0.135}  \\
\midrule
Ref.~\cite{Gaines1974}       &   12\,178\,595\phantom{.000}  & & & 3\phantom{.000}  \\
Ref.~\cite{Breckenridge1995} &   12\,178\,595\phantom{.000}  & & & 4\phantom{.000}  \\
Ref.~\cite{Lovas2004}        &   12\,178\,593\phantom{.000}  & & & 4\phantom{.000}  \\
Ref.~\cite{Coudert2015}      &   12\,178\,600\phantom{.000}  & & & 2\phantom{.000}  \\
Ref.~\cite{Collignon2026}    &   12\,178\,596.106  & & & 0.243  \\
\bottomrule
\end{tabular}
\end{table}

With the absolute frequencies and systematic error budgets of the six targeted optical lines established in Sec.~\ref{sec:results}, we now turn to the extraction of the $3_{-1}$E -- $2_0$E microwave transition frequency.
The determination of the $3_{-1}$E -- $2_0$E microwave transition frequency relies on the \emph{difference} between two optical transitions ($\Delta_{i,1}$ and $\Delta_{i,2}$). Because the optical frequencies of the six transitions differ by less than $0.01\%$, the SODS acts as a common-mode systematic effect that cancels entirely in the subtraction $\Delta_i = \nu(\Delta_{i,1}) - \nu(\Delta_{i,2})$. 
In contrast, the centroid shift associated with unresolved hyperfine structure comes in twice for which we take a combined systematic boundary ($\sigma_{\text{sys,}\Delta}  \approx 10$~Hz), which is then added linearly to $\sigma_{\text{stat}}$.

\begin{figure}[htbp]
\includegraphics[width=1\linewidth]{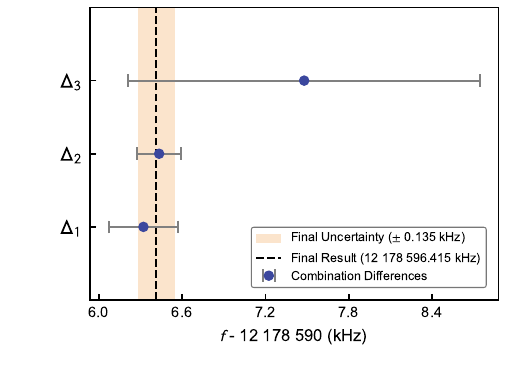}
\caption{Final determination of the $3_{-1}$E -- $2_{0}$E rotational ground state frequency. The three independent values derived from the combination differences ($\Delta_1$, $\Delta_2$, and $\Delta_3$) are plotted as data points with their $1\sigma$ propagated uncertainties(from Table~\ref{tab:radio}). The agreement between the high-precision measurements $\Delta_1$ and $\Delta_2$ is evident. The $\Delta_3$ measurement, which had a larger uncertainty due to the spectral contamination discussed in Sec.~\ref{sec:neighboring_resonance}, is fully consistent with the other two. 
The final weighted mean of $12\,178\,596.415~\text{kHz}$ is shown as a dashed vertical line, with the orange shaded band representing its $1\sigma$ standard error of $0.135~\text{kHz}$.}
\label{fig:final}
\end{figure}

The resulting values for the radio line, calculated via the combination differences displayed in Fig.~\ref{fig:triangle}, are compiled in Table~\ref{tab:radio} and plotted in Fig.~\ref{fig:final}. The final derived frequency is taken as the weighted mean of the three independent determinations ($\Delta_1$, $\Delta_2$, $\Delta_3$), calculated using purely statistical weights (derived strictly from $\sigma_{\text{stat}}$).

This final aggregated value represents the central centroid of the microwave transition. For astrophysical comparisons, this centroid is precisely the relevant measure. Because the lines observed in radio astronomy are typically thermally broadened beyond the composite hyperfine structure, the underlying multiplet effectively blends into a single symmetric profile. This holds for the extra-galactically observed methanol lines, where the $3_{-1}$E -- $2_{0}$E line was observed at a width of 750 kHz~\cite{Bagdonaite2013a,Bagdonaite2013b,Kanekar2015}. Even the much narrower lines observed in cold cores in our Galaxy (at widths in the range 50 - 80 kHz) appear with fully symmetric line profiles~\cite{Dapra2017b}. This does not hold for studies based on methanol maser lines that typically operate on selected hyperfine components~\cite{Agafonova2024}.

Several independent measurements of the $3_{-1}$E -- $2_0$E radio line have been reported over the past decades. Kukolich and coworkers~\cite{Gaines1974,Breckenridge1995} published two values with accuracies of 3 and 4 kHz, respectively; these are in excellent agreement with the present, more precise determination. The result is also consistent with the NIST database value~\cite{Lovas2004}, which is derived from a global fit to a comprehensive dataset of methanol rotational transitions~\cite{Xu2008}.
Coudert et al.~\cite{Coudert2015} performed partially hyperfine-resolved microwave measurements on methanol, yielding a value for the centroid frequency accurate to 2 kHz.
Finally, a concurrent measurement was performed at UCLouvain using the technique of Free-Induction-Decay (FID)~\cite{Collignon2026}. That study yields a transition frequency with an uncertainty of 243 Hz, only slightly less accurate than achieved here. The difference between the two determinations is just 309 Hz (corresponding to $1.1\sigma$ for the combined uncertainties) obtained using entirely independent methods supports the validity of the triangulation scheme applied here for internal rotor molecules. 

\section{Conclusion}

In the present near-infrared saturation study of methanol (CH$_3$OH), the transition frequencies of rovibrational lines are measured with absolute accuracies as low as $120$~Hz, representing a  new landmark in high-resolution spectroscopy of molecules.
A variety of modern cavity-enhanced frequency-comb locked molecular metrology studies have been performed, with only a few achieving sub-kHz accuracy~\cite{Wang2017,Reed2020,Tan2022}.
Specifically for methanol, an improvement of almost two orders of magnitude in spectroscopic accuracy is achieved over the most accurate previous study, which was performed at much lower frequencies (29 THz)~\cite{Santagata2019}, and which in itself was over three orders of magnitude more accurate than  Fourier-transform infrared techniques~\cite{Xu2004a}.

By analyzing combination differences between transitions measured at 216 THz, the transition frequency of the important 12.2 GHz methanol line was determined to an uncertainty of 135 Hz, surpassing the precision obtained in previous direct microwave measurements by a factor of $\sim$20. This $3_{-1}$E -- $2_0$E rotational transition in the vibrationless ground state of methanol is one of the most sensitive spectral lines for detecting a possible variation of the proton-to-electron mass ratio on cosmological time scales (with a large sensitivity coefficient of $K_{\mu}=-33$), underscoring its significance. We report a value of $12\,178\,596.415$(135)~kHz for this prominent radio astronomical benchmark line.
 
The accurate frequency of this microwave transition is derived from a novel triangulation scheme that is fundamentally different from the schemes found in chiral molecules or symmetric top rotors. Here, a closed loop of exactly three electric-dipole transitions is made possible because the torsion-rotation states of $E$-symmetry in methanol exhibit mixed parity. This methodology can be applied to extract frequencies of other E-type radio lines in methanol.

\section*{Backmatter}

\subsection*{Funding}
The research leading to these results has received funding from LASERLAB-EUROPE (grant agreement no. 871124, European Union’s Horizon 2020 research and innovation programme).
The present study is also part of a European Partnership on Metrology project, which is co-financed from the European Union’s Horizon Europe Research and Innovation Programme and by the Participating States (Funder ID: 10.13039/100019599, Grant number: 23FUN04 COMOMET).
WU acknowledges funding via an ERC-Advanced grant under the European Union’s Horizon 2020 research and innovation programme (Grant Agreement No. 670168).
The work at Louvain was supported by the Fonds de la Recherche Scientifique – FNRS (CDR grant no. J.0124.26 and MIS grant no. F.4038.24). 
FMJC is a Chargé de recherche of the Fonds de la Recherche Scientifique – FNRS.
ASB is supported by a Marie Curie fellowship provided by the EU Horizon 2023 program. 
IK thanks the Agence Nationale de la recherche Scientifique (ANR) project ANR-25-CE29-7779. BL acknowledges funding from the European Union’s Horizon Europe research and innovation programme under the Marie Skłodowska-Curie grant agreement No. 101126636.
The work of JCJK was supported by the Dutch National Growth Fund, as part of the Quantum Delta NL programme.

\subsection*{Acknowledgment}

The setup was built with help of the mechanical and electronic workshops at VU University Amsterdam.
The Van Swinden Laboratory (Netherlands Metrology Institute) is acknowledged for the use of a fiber link to their frequency standards. 
This work was made possible through the Netherlands SURF Time \& Frequency Network.
The authors thank the anonymous referees for their suggestions that have led to improvement of this manuscript.

\subsection*{Disclosures}
The authors declare no conflicts of interest.

\subsection*{Supplementary Material}
Several topics presented in this manuscript are described in more detail: 
(i) Quantitative effects of hyperfine substructure on centroid frequency of the $3_{-1}$E -- $2_{0}$E transition. (ii) Generalized approach to centroid shifts as a result of underlying hyperfine structure. 
(iii) Estimation of systematic uncertainty.
(iv) Long-term reproducibility of the measurement setup. (v) Analysis of linewidth including saturation effects.

\subsection*{Data Availability}
Data underlying the results presented in this paper 
may be obtained from the authors upon reasonable request.


\bibliography{NICE-OHMS-2025,Var-Const,Hydrogen-2025,Lasers+Spectroscopy}

\end{document}